\documentclass[onecolumn,showpacs,preprintnumbers,amsmath,amssymb,12pt]{revtex4}

\usepackage{graphicx}
\usepackage{dcolumn}
\usepackage{bm}
\begin{document}

\preprint{APS/123-QED}

\title{Calculated magnetoresistance due to domain walls in nanostructures}

\author{M. C. Hickey}
\affiliation{%
School of Physics and Astronomy, E. C. Stoner Laboratory, University
of Leeds, Leeds, LS2 9JT, United Kingdom.
}%


\date{\today}

\begin{abstract}
  The existing Levy-Zhang approach to constructing the contribution to the
resistivity of a metal of a magnetic domain wall is explored. The model equations are
integrated analytically, giving a closed form expression for the resistivity
when the current flows in the wall. The Boltzmann equation is solved
analytically and the ratio of the spin up and spin down resistivities is
calculated and its dependence on the strength of the Coulomb and exchange
scattering potentials is elucidated.
\end{abstract}

\maketitle

\section{\label{sec:level1}Introduction}
Domain walls are examples of topological solitons in magnetism and
they arise due to the competition between exchange and anisotropy
energy. Domain wall motion by a spin polarized current has been
gathering much interest recently mainly due to emerging device
applications such as domain wall memory and domain wall logic
devices. Winding number (vorticity), chirality and even skyrmion
number are other degrees of freedom when considering the magnetic
domain wall and this is fascinating from the point of view of
fundamentals as well as information storage considerations.
Understanding the mechanisms by which a magnetic domain wall
contributes to the resistivity of a metal, is a problem on equal
footing with that of describing how a spin polarized current imparts
torque to magnetization. When a conduction band electron fails to
track the lattice magnetization when traversing a domain wall, an
angle is subtended between the conduction band spin and the wall,
which leads to a torque and, in the presence of impurity scattering,
a measurable magnetoresistance. The relationship between domain wall
motion spin transfer torque and domain wall magnetoresistance
was proposed by Tatara {\it et al.} \cite{PhysRevLett.78.3773}.\\
In this paper, we are interested in calculating analytically the
explicit formula for the contribution of a domain wall to the
resistivity in the diffusive limit, using the model equations of
Levy and Zhang \cite{PhysRevLett.79.5110}. We wish to integrate this
model, giving explicit formulae for the resistivity of a domain wall
and use this formalism to calculate MR curves for systems in which
domain walls nucleate in nanostructures by shape anisotropy.

\section{\label{sec:level2}Admixture states at a domain wall}
\begin{figure}[h]
\begin{center}
 \includegraphics[width=4.0in]{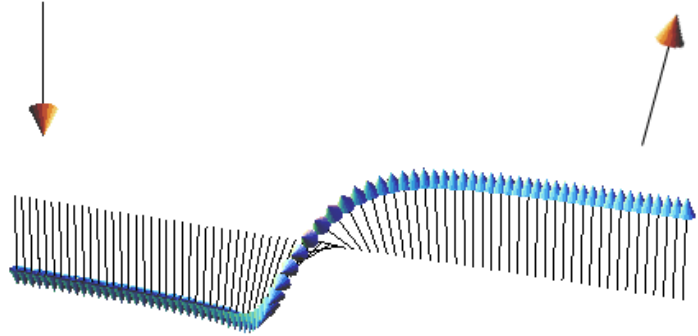}\\
\end{center}
 \caption{(Color Online) Schematic of a conduction band electron traversing a 180$^{o}$
 Bloch domain wall and undergoing mistracking.} \label{fig1}
\end{figure}

We first begin with the simple picture of a 2 band ferromagnetic
metal where the Fermi level lies in the Stoner split bands. We
consider the Hamiltonian of a uniformly magnetized ferromagnet with
the unit vector of magnetization aligned along the +z axis
($\sigma.\hat{n} = \sigma_{z}$), and so the starting SU(2)
Hamiltonian takes the following form :

\begin{equation}
H_{0}=\frac{-\hbar^{2}\nabla^{2} }{2 m^{*}}+ V(\overrightarrow{r})+J
\sigma_{z}, \label{H0}
\end{equation}
where m$^{*}$ is the effective electron mass and
V($\overrightarrow{r}$) is the periodic crystal potential, taken to
be invariant under SU(2) rotation and therefor this does not
contribute to the spin scattering in the analysis which follows. We
now write the Hamiltonian H$_{0}$ in matrix form and look for
eigenstates in the Hilbert space $L_{2}\otimes{\emph H_{s}}$.

\begin{equation}
H_{0} = \left(
          \begin{array}{cc}
            -\beta \nabla^{2}+J & 0 \\
            0 & \beta \nabla^{2}-J \\
          \end{array}
        \right),
\end{equation}
where $\beta$ =$\hbar^{2}/(2 m^{*})$. We now transform H$_{0}$ onto
the basis $\{e^{i
\overrightarrow{k_{\sigma}}.\overrightarrow{r}}\}$, which assumes
eigenvectors of the form $\Phi_{\overrightarrow{k} \sigma}$=$e^{i
\overrightarrow{k_{\sigma}}.\overrightarrow{r}}$ $\varphi_{s}$,
where $\varphi_{s}$ is a two-component spinor. Writing
$\hat{H}=\langle\Phi^{*}_{\overrightarrow{k}
\sigma}|H_{0}|\Psi_{\overrightarrow{k}\sigma}\rangle$, we find :

\begin{equation}
\hat{H} = \left(
            \begin{array}{cc}
              \beta k_{\sigma}^{2}+J & 0 \\
              0 & -\beta k_{\sigma}^{2}-J \\
            \end{array}
          \right).
          \label{eigneveq}
\end{equation}
The eigenvalues of  Equation \ref{eigneveq} can be written as :
$\lambda_{\pm}$ =$\beta k^{2}_{\pm}\pm J$ where the $\pm$ signs
refer to pure spin eigenstates.
We now write the 2 component spinors for an
unperturbed 2 band, exchange split ferromagnet.
\begin{equation}
\phi^{(0)}_{\uparrow}=\frac{1}{\sqrt{N}}\left(
                                           \begin{array}{c}
                                             e^{i\overrightarrow{k_{\uparrow}}.\overrightarrow{r}} \\
                                             0 \\
                                           \end{array}
                                         \right),\\
\phi^{(0)}_{\downarrow}=\frac{1}{\sqrt{N}}\left(
                                           \begin{array}{c}
                                             0 \\
                                             e^{i\overrightarrow{k_{\downarrow}}.\overrightarrow{r}} \\
                                           \end{array}
                                         \right),
\end{equation}
which describe pure spin states characterizing a two-band
ferromagnet each with eigen energies E$_{\uparrow,\downarrow}$ =
$\beta$ k$_{\uparrow,\downarrow}^{2}\pm J$. We now turn our
attention to the perturbation associated with a magnetic domain wall
(DW), where the description of the conduction band spin goes beyond
that of pure spins states. If the region of space over which the
magnetization in a DW rotates is comparable the length scale of the
Fermi wavelength 1/k$_{F}$, there will be an adiabatic 'mistracking'
of the conduction spin with the lattice magnetization (see Fig.
\ref{fig1}). Nevertheless, as we will see, this scaling is treated as
a perturbation in the Levy-Zhang approach. The perturbation
parameter is proportional to $\nabla \theta.\overrightarrow{k_{F}}$
and this is assumed to be small for the perturbation expansion to
converge.
We write the Hamiltonian defined in equation \ref{H0} in the
transformed basis of the domain wall R$_{\theta}\Phi^{(0)}_{\sigma}$
which gives :
\begin{equation}
\bar{H} = R^{-1}_{\theta}H R_{\theta} =
H_{0}+R^{-1}_{\theta}[H,R_{\theta}],
\end{equation}
where H$_{0}$ is now the unperturbed Hamiltonian of the magnetic
system and R$_{\theta}$ is the SU(2) rotation operator e$^{-i\frac{
\theta}{2} \hat{n}.\sigma}$. Indeed, in the basis of pure spin
states, $\theta$ is the polar angle of the magnetization unit vector
$\hat{n}$. We recognize that the perturbation potential can be
written from the equation above as
$V_{pert}$=$R^{-1}_{\theta}[H,R_{\theta}]$. Now, R$_{\theta}$
commutes with the J $\hat{n}.\sigma$  and
V($\overrightarrow{r}$)terms in H$_0$, so we are left with
V$_{pert}$ in the following form :
\begin{eqnarray}
V_{pert} = -\beta R^{-1}_{\theta}[\nabla^{2},R_{\theta}]\\
=-\beta R^{-1}_{\theta}(\nabla^{2}R_{\theta} -
R_{\theta}\nabla^{2})\\
=-\beta (R^{-1}_{\theta}\nabla^{2}R_{\theta}-\nabla^{2}).
\label{Vpert}
\end{eqnarray}
Now, in order to evaluate the left hand side term of the above
equation $R^{-1}_{\theta}\nabla R_{\theta}$, we act on a trial
wavefunction $\psi$ from the left as follows :
\begin{eqnarray}
(R^{-1}_{\theta}\nabla.\nabla R_{\theta}) \psi\\
=R^{-1}_{\theta}\nabla(\nabla R_{\theta}\psi+R_{\theta}\nabla\psi)\\
=R^{-1}_{\theta}(\nabla^{2}R_{\theta}+(\nabla R_{\theta})\nabla
\psi+(\nabla R_{\theta})\nabla \psi+R_{\theta}\nabla^{2}\psi)\\
=R^{-1}_{\theta}(\nabla^{2}R_{\theta}\psi+(\nabla R_{\theta})\nabla
\psi+(\nabla R_{\theta})\nabla \psi+R_{\theta}\nabla^{2}\psi)\\
=(R^{-1}_{\theta}\nabla^{2}R_{\theta}+2R^{-1}_{\theta}\nabla
R_{\theta}\nabla+\nabla^{2})\psi)\\
\end{eqnarray}
Inserting this into equation \ref{Vpert}, we arrive at the following
expression ;
\begin{equation}
V_{pert} =
-\beta(R^{-1}_{\theta}\nabla^{2}R_{\theta}-\nabla^{2})\psi\\
=-\beta(R^{-1}_{\theta}\nabla^{2}R_{\theta}+2R^{-1}_{\theta}(\nabla
R_{\theta})\nabla+\nabla^{2})\psi
\end{equation}
We can can simply write $\nabla R_{\theta} =-i
\frac{\nabla\theta}{2}(\hat{n}.\sigma) e^{-i
\frac{\theta}{2}\hat{n}.\sigma}$ and $\nabla^{2}
R_{\theta}=-\frac{|\nabla\theta|^{2}}{4}(\hat{n}.\sigma)^{2}e^{-i
\frac{\theta}{2}\hat{n}.\sigma}$ -
$i\frac{\nabla^{2}\theta}{2}(\hat{n}.\sigma) e^{-i
\frac{\theta}{2}\hat{n}.\sigma}$ which, when substituted into
V$_{pert}$ now gives :
\begin{equation}
V_{p}=-\beta\left[-\frac{|\nabla\theta|^{2}}{4}(\hat{n}.\sigma)^2-i\frac{\nabla^{2}\theta}{2}(\hat{n}.\sigma)-i\nabla\theta(\hat{n}.\sigma).\nabla\right]
\end{equation}
Recall that $\beta$ = $\hbar^{2}/2 m^{*}$ which is just a constant
and that $\theta$ is the angle of the magnetization. Recognizing
that, for generators of SU(2) rotations, $(\hat{n}.\sigma)^{2}$ = 1,
which means that the first term in V$_{pert}$ is diagonal and so
does not mix spin states. Further, if the wall magnetization is
assumed to be slowly varying in space with respect to the length
scale defined by $1/k_{F}$, we have $\nabla^{2}\theta/(2 \nabla
\theta k_{F})$ $\ll$ 1. This latter term may become important in DW
profiles with vanishing $\nabla\theta$ but finite $\nabla^{2}\theta$
(i.e. a stationary point in $\theta$) which would occur in DW
configurations with finite winding (n$\geq$1) or skyrmion number. As a
first approximation, we retain the first order term in V$_{pert}$ =
-$\beta \nabla \theta(\hat{n}.\sigma)(-i\nabla)$ and use the
perturbation formalism outlined in Appendix A. We write the 
new eigenspinors in the rotated basis as :

\begin{eqnarray}
|\uparrow^{'}\rangle = R_{\theta}\left(
                               \begin{array}{c}
                                 e^{i\overrightarrow{k}_{\uparrow}.\overrightarrow{r}} \\
                                 0 \\
                               \end{array}
                             \right)\\
|\downarrow^{'}\rangle = R_{\theta}\left(
                               \begin{array}{c}
                                 0 \\
                                 e^{i\overrightarrow{k}_{\downarrow}.\overrightarrow{r}} \\
                               \end{array}
                             \right)
\end{eqnarray}
For a Bloch wall, where the magnetization rotates in the yz plane,
we now write the expansion coefficients for the first order
corrections to the wavefunction (see Appendix A), as follows  :
\begin{eqnarray}
C^{(1)}_{k} =
\displaystyle\sum_{n\neq k}\frac{V_{nk}}{E^{(0)}_{n}-E^{(0)}_{k}}\\
=\displaystyle\sum_{n\neq k}\frac{\langle k^{(0)}|-\beta \nabla
\theta.-i\nabla(\hat{n}.\sigma)|n^{(0)}\rangle}{E^{(0)}_{n}-E^{(0)}_{k}}\\
= C^{(1)}_{\uparrow} =\frac{\langle\uparrow|-\beta(\nabla
\theta).(-i\nabla
\theta)(\hat{n}.\sigma)|\downarrow\rangle}{E^{(0)}_{\downarrow}-E^{(0)}_{\uparrow}}\\
=\frac{\int d^{3}\overrightarrow{r}(-i\beta
\nabla\theta.\overrightarrow{k_{\downarrow}}e^{i(
\overrightarrow{k}_{\uparrow}-\overrightarrow{k}_{\downarrow}).\overrightarrow{r}}
)}{-2 J  +\beta (k_{\downarrow}^{2}-k_{\uparrow}^{2})}
\end{eqnarray}
We find a similar expression for the C$^{(1)}_{\downarrow}$ mixing
coefficient. It is important to note also that the unit vector along
the magnetization can be written as $\hat{n} =
(0,\sin(\theta)\sin(\phi),\cos(\theta))
=(0,sech(x/\lambda)\sin(\phi),\tanh(x/\lambda))$, for a Bloch wall
in the +x direction with chirality $\phi$. $\lambda$ is the
equilibrium wall width (=$\sqrt(A/K)$, A being the exchange
stiffness and K is the magnetic anisotropy energy density). In this
wall configuration, $\hat{n}.\sigma$ has the components
$n_{y}\sigma_{2}+n_{z}\sigma_{3}$, where $\sigma_{i}$ refers to the
components of the Pauli spinors. Moreover, only the $\sigma_{2}$
term yields a non-zero contribution to the mixing coefficient as
it's elements are off diagonal. If the wall is set up to rotate in
the xz plane, the coefficient C$^{(1)}$ would be real.

\begin{eqnarray}
C^{(1)}_{\uparrow}= \frac{\int d^{3}\overrightarrow{r}(i\beta( e^{i(
-\overrightarrow{k}_{\uparrow}+\overrightarrow{k}_{\downarrow}).\overrightarrow{r}}
\nabla
\theta.\overrightarrow{k_{\downarrow}}+\frac{\nabla^{2}\theta}{2}e^{i(
-\overrightarrow{k}_{\uparrow}+\overrightarrow{k}_{\downarrow}).\overrightarrow{r}}))}{\beta
(k^{2}_{\downarrow}-k^{2}_{\uparrow})-2 J}\\
C^{(1)}_{\downarrow}= \frac{\int d^{3}\overrightarrow{r}(-i\beta(
e^{i(
-\overrightarrow{k}_{\downarrow}+\overrightarrow{k}_{\uparrow}).\overrightarrow{r}}
\nabla
\theta.\overrightarrow{k_{\uparrow}}+\frac{\nabla^{2}\theta}{2}e^{i(
-\overrightarrow{k}_{\downarrow}+\overrightarrow{k}_{\uparrow}).\overrightarrow{r}}))}{\beta
(k^{2}_{\uparrow}-k^{2}_{\downarrow})+2 J}\\
\end{eqnarray}
In ferromagnetic metals, it is reasonable to assume that the kinetic
energy splitting between the bands is much smaller than the exchange
splitting , this condition is written as $\frac{\beta
(k^{2}_{\uparrow}-k^{2}_{\downarrow})}{2 J} \ll 1$. It can further
be assumed that  $k^{F}_{\uparrow}\simeq k^{F}_{\downarrow}$, which
is true to within an order of magnitude for most ferromagnets, and
for the purposes of this calculation, the assumption is convenient
in establishing the order of of magnitude of the effect. For
spatially dependent $\nabla\theta$, $\nabla^{2}\theta$ and
$\hat{n}(\overrightarrow{r}).\sigma$ this assumption would have to
be relaxed and these terms will couple to the scattering
coefficients $C^{(1)}_{\uparrow,\downarrow}$ via a transformation on
the basis kets $\{|\uparrow^{'}\rangle,|\downarrow^{'}\rangle\}$ via
:
\begin{eqnarray}
\langle\uparrow^{'}|i\beta
\nabla\theta.\overrightarrow{k}_{\downarrow}\hat{n}(\overrightarrow{r}).\sigma\nabla|\downarrow^{'}\rangle,\\
\langle\downarrow^{'}|-i\beta
\nabla\theta.\overrightarrow{k}_{\uparrow}\hat{n}(\overrightarrow{r}).\sigma\nabla|\uparrow^{'}\rangle,
\end{eqnarray}
which correspond to the C$^{(1)}_{\uparrow}$ and
C$^{(1)}_{\downarrow}$ mixing coefficients, respectively. We now
write the total wavefunction of the electron in terms of the
adiabatically mixed two
component spinors in the rotated basis, as follows : \\
\begin{eqnarray}
|\Psi_{\uparrow}^{'}\rangle  =
\frac{1}{\sqrt{N}}\left[R_{\theta}\left(
                                                                \begin{array}{c}
                                                                  e^{i\overrightarrow{k}_{\uparrow}.\overrightarrow{r}} \\
                                                                  0 \\
                                                                \end{array}
                                                              \right)
+C^{(1)}_{\uparrow} R_{\theta}\left(
                                \begin{array}{c}
                                  0 \\
                                  e^{i \overrightarrow{k}_{\uparrow}.\overrightarrow{r}} \\
                                \end{array}
                              \right)\right]\\
|\Psi_{\downarrow}^{'}\rangle  =
\frac{1}{\sqrt{N}}\left[R_{\theta}\left(
                                                                \begin{array}{c}
                                                                  0 \\
                                                                  e^{i\overrightarrow{k}_{\downarrow}.\overrightarrow{r}} \\
                                                                \end{array}
                                                              \right)
+C^{(1)}_{\downarrow} R_{\theta}\left(
                                \begin{array}{c}
                                  e^{i \overrightarrow{k}_{\downarrow}.\overrightarrow{r}}  \\
                                  0\\
                                \end{array}
                              \right)\right]
\end{eqnarray}
Using the approximations implemented by Levy and Zhang, we can write
$C^{(1)}_{\uparrow} = -i\frac{k_{x}\xi}{k_{F}}$ and
$C^{(1)}_{\downarrow} =i \frac{k_{x}\xi}{k_{F}}$  for a wall whose
magnetization rotates along the x-axis in the adopted coordinate
system. We now define the 'spin mistracking' parameter as :
\begin{equation}
 \xi  = \frac{\beta \nabla\theta.\overrightarrow{k_{F}}}{2 J}
\end{equation}
Here $\nabla\theta$ is the local magnetization angle gradient, which
can be taken to be locally constant. The normalization coefficients,
can be written as follows ;
$N_{\sigma}=\|\Psi^{\dag}_{\sigma}\Psi_{\sigma}\| = 1$, we find that
N(k$_{x}$)= $1+(\frac{k_{x}\xi}{k_{F}})^2$. The value of
$\nabla\theta$ can be taken to be locally constant over the
lengthscale 1/k$_{F}$ and for linearly varying magnetization
profiles, the mistracking can be written as

\begin{equation}
 \xi  = \frac{\hbar^{2}k_{F}\pi}{4mJ D}
\end{equation}
which describes mistracking at the wall whose profile is - $\theta(x) =
\frac{\pi x}{D}$  (-D$<$x$<$D), originally considered by Levy and
Zhang. We now write the corrections to the total wavefunction to
first order as

\begin{eqnarray}
|\Psi_{\uparrow}^{'}(\overrightarrow{k},\overrightarrow{r})\rangle =
\frac{1}{1+(\frac{\xi k_{x}}{k_{F}})^{2}}\left[R_{\theta}\left(
                                                                \begin{array}{c}
                                                                  e^{i\overrightarrow{k}_{\uparrow}.\overrightarrow{r}} \\
                                                                  i\frac{k_{x}\xi}{k_{F}}e^{i \overrightarrow{k}_{\uparrow}.\overrightarrow{r}} \\
                                                                \end{array}
                                                              \right)\right]\\
|\Psi_{\downarrow}^{'}(\overrightarrow{k},\overrightarrow{r})\rangle
= \frac{1}{1+(\frac{\xi k_{x}}{k_{F}})^{2}}\left[R_{\theta}\left(
                                                                \begin{array}{c}
                                                                  -i\frac{k_{x}\xi}{k_{F}}e^{i \overrightarrow{k}_{\downarrow}.\overrightarrow{r}} \\
                                                                  e^{i\overrightarrow{k}_{\downarrow}.\overrightarrow{r}} \\
                                                                \end{array}
                                                              \right)\right]
                                                              \label{rotated_basis}
\end{eqnarray}

\section{\label{sec:level3}Impurity scattering at a domain wall}
The domain wall itself does not necessarily give rise to inelastic scattering or
to a measurable resistance. However, let us consider what happens
when we consider an impurity potential to which the conduction spin
is coupled via the coulomb and exchange interaction. The scattering
potential is defined as follows :
\begin{equation}
V_{scatt}(\overrightarrow{r}) =
\displaystyle\sum_{i}\left[v+j\sigma.\hat{M}(\overrightarrow{r}_{i})\right]\delta(\overrightarrow{r}-\overrightarrow{r}_{i})
\end{equation}

This scattering potential has the following matrix elements in the
basis $\{|\Psi_{\sigma}(\overrightarrow{r})\rangle\}$ as follows :
\begin{equation}
V^{\sigma\sigma^{'}}_{kk^{'}}=\displaystyle\int
d^{3}\overrightarrow{r}\Psi^{\dag}_{\sigma}(\overrightarrow{k},\overrightarrow{r})V_{scatt}\Psi_{\sigma^{'}}(\overrightarrow{k'},\overrightarrow{r}).
\end{equation}
Using the basis defined by equation \ref{rotated_basis}, we write
down the matrix elements of the scattering potential :
\begin{eqnarray}
V^{\uparrow\uparrow}_{kk^{'}}=\displaystyle\sum_{i}\int\frac{1}{N(k)N(k^{'})}R^{-1}_{\theta}\left(
                                                                                               \begin{array}{cc}
                                                                                                 e^{-i{\bf k}_{\uparrow}.\overrightarrow{r}} & C^{*(1)}_{\uparrow}e^{-i{\bf k}_{\uparrow}.\overrightarrow{r}} \\
                                                                                               \end{array}\right).\\
                                                                                               \left[R_{\theta}v\left(\begin{array}{c}
                                                                                                                             e^{i{\bf k}^{'}_{\uparrow}.\overrightarrow{r}} \\
                                                                                                                             C^{(1)}_{\uparrow}e^{i{\bf k}^{'}_{\uparrow}.\overrightarrow{r}}\\
                                                                                                                           \end{array}
                                                                                             \right)+j\sigma.\hat{M}(\overrightarrow{r}_{i})R_{\theta}\left(
                                                                                                                                             \begin{array}{c}
                                                                                                                                               e^{i{\bf k}^{'}_{\uparrow}.\overrightarrow{r}} \\
                                                                                                                                               C^{(1)}_{\uparrow}e^{i{\bf k}^{'}_{\uparrow}.\overrightarrow{r}} \\
                                                                                                                                         \end{array}\right)\right]\delta(\overrightarrow{r}-\overrightarrow{r}_{i})d^{3}\overrightarrow{r}
\end{eqnarray}

Note : $|\displaystyle\sum_{i} e^{i({\bf k^{'}}-{\bf k}).r_{i}}|^{2}=\displaystyle\sum_{i,j}e^{i({\bf k}^{'}-{\bf k}).{\bf r}_{i}}e^{i({\bf k}^{'}-{\bf k}).{\bf r_{j}}} = \displaystyle\sum_{i}1+ \displaystyle\sum_{i\neq j}e^{i({\bf k}^{'}-{\bf k}).({\bf r_{i}}- {\bf r}_{j})}= c_{i}\Omega$
,where c$_{i}$ refers to the concentration of impurity scattering
sites. This counting is an average over all of the impurity sites and each site is
 taken to be equivalent. Having established the matrix elements of the scattering
potential, we write down the scattering rates based on Fermi's
golden rule. After integration, the scattering matrix
elements can be written in the following form :
\begin{eqnarray}
\|V^{\uparrow\uparrow}_{kk^{'}}\|^{2}=c_{i}\frac{1}{N^{2}(k_{x})N^{2}(k^{'}_{x})}\left[(v+\sigma
j)+\frac{k_{x} k^{'}_{x}}{k^{2}_{F}}\xi^{2}(v-\sigma
j)\right]^{2}\\
\|V^{\uparrow\downarrow}_{kk^{'}}\|^{2}=c_{i}\frac{1}{N^{2}(k_{x})N^{2}(k^{'}_{x})}\left[(v+j)k^{'}_{x}-(v-j)k_{x}\right]^{2}\frac{\xi^{2}}{k^{2}_{F}}
\end{eqnarray}

\begin{equation}
W^{\sigma\sigma^{'}}_{kk^{'}} =
\frac{2\pi}{\hbar}\delta(\epsilon_{{\bf k}\sigma}-\epsilon_{{\bf
k}^{'}\sigma})\|V^{\sigma\sigma^{'}}_{kk^{'}}\|^{2} .
\end{equation}
\begin{eqnarray}
W^{\sigma\sigma}_{{\bf kk^{'}}} = \frac{2
\pi}{\hbar}c_{i}\frac{1}{N^{2}(k_{x})N^{2}(k^{'}_{x})}\left[(v+\sigma
j)+\frac{k_{x} k^{'}_{x}}{k^{2}_{F}}\xi^{2}(v-\sigma
j)\right]^{2}\\
W^{\sigma\sigma^{'}}_{kk^{'}}=\frac{2
\pi}{\hbar}c_{i}\frac{1}{N^{2}(k_{x})N^{2}(k^{'}_{x})}\left[(v+j)k^{'}_{x}-(v-j)k_{x}\right]^{2}\frac{\xi^{2}}{k^{2}_{F}}
\label{scattering_rates}
\end{eqnarray}
These scattering rates can be integrated over momentum space (${\bf
k}^{'}$ coordinates) in order to find the total scattering lifetimes
for momentum states within a spin channel and for momentum
scattering which mixes the spin channels. This total scattering rate is defined as
follows :

\begin{equation}
[\tau^{\sigma}({\bf k})]^{-1} =
\frac{1}{(2\pi)^{3}}\displaystyle\int d^{3}{\bf
k}^{'}(W^{\sigma\sigma}_{{\bf kk}^{'}}+W^{\sigma\sigma}_{{\bf
kk}^{'}})
\end{equation}
\begin{eqnarray*}
[\tau^{\sigma}({\bf k})]^{-1} =\frac{1}{(2\pi)^3}\displaystyle\int
d^{3}{\bf
k}N^{-2}(k_{x})N^{-2}(k^{'}_{x})\\
\left[((v+j)k^{'}_{x}-(v-j)k_{x})^{2}\frac{\xi}{k^{2}_{F}}
+((v+\sigma j)+\frac{k_{x}k^{'}_{x}}{k^{2}_{F}}\xi^{2}(v-\sigma
j))\right]^{2}\delta(\epsilon_{{\bf k}\sigma} - \epsilon_{{\bf
k}^{'}\sigma}). \label{tau_int}
\end{eqnarray*}
We expand the normalization constants $N(k_{x})$ to second order
$\xi$ as follows ; $N^{-2}(k_{x})\simeq
1-(\frac{k_{x}\xi}{k_{F}})^2+O((\frac{k_{x}\xi}{k_{F}})^{4})$ and we
apply this approximation in order to evaluate the integral in
Equation \ref{tau_int}. Recognizing that Equation \ref{tau_int} has
integrals of two types, we define these two types as follows ;

\begin{eqnarray}
I_{1}=\frac{2\pi c_{i}}{\hbar}\displaystyle\int
d^{'}k^{3}\frac{\xi^{2}}{k^{2}_{F}}(A
k^{'}_{x}+B)^{2}(1-2\frac{(k_{x}k^{'}_{x})^{2}}{k^{2}_{F}}\xi^{2}+..)\delta(\epsilon_{{\bf
k}\sigma}-\epsilon_{{\bf k}^{'}\sigma})\\
I_{2}\frac{2\pi c_{i}}{\hbar}\displaystyle\int
d^{'}k^{3}\frac{\xi^{2}}{k^{2}_{F}}(C +D
\frac{k^{'}_{x}\xi^{2}}{k^{2}_{F}})^{2}(1-2\frac{(k_{x}k^{'}_{x})^{2}}{k^{2}_{F}}\xi^{2}+..)\delta(\epsilon_{{\bf
k}\sigma}-\epsilon_{{\bf k}^{'}\sigma}),\\
\end{eqnarray}
where we define the following constants within the integral.
\begin{displaymath}
A = (v+j),\\
B=-(v-j)k_{x},\\
C=(v+\sigma j),\\
D=k_{x}(v-\sigma j).
\end{displaymath}
The {\bf k} space volume element is given in spherical polar
coordinates as  $d^{3} {\bf k}$ = $k^{2}\sin(\theta)\cos(\phi)dk$
and we write down \textit{I$_{1}$} using this coordinate system,
while expanding to second order in $\xi$, as follows :
\begin{eqnarray}
\begin{array}{c}
  I_{1}=\frac{2\pi c_{i}}{\hbar}\displaystyle\int d^{'}k
k^{'2}\sin(\theta)d\phi d\theta\frac{\xi^{2}}{k^{2}_{F}}(A
k^{'}\sin\theta^{'}\cos\phi^{'}+B)^{2}\times \\
  (1-2k^{2}_{x}k^{'2}\sin^{2}\theta^{'}\cos^{2}\phi^{'}\frac{\xi^{2}}{k^{2}_{F}}+..)\delta(\epsilon_{{\bf k}\sigma} -\epsilon_{{\bf k}^{'}\sigma}). \\
  \displaystyle\int d^{'}k k^{'2}\sin(\theta)d\phi
d\theta\frac{\xi^{2}}{k^{2}_{F}}(A^{2}k^{'2}\sin^{2}\theta^{'}\cos^{2}\phi^{'}+2ABk^{'}\sin\theta^{'}\cos\phi^{'}+B^{2})\times\\
 (1-2k^{2}_{x}k^{'2}\sin^{2}\theta^{'}\cos^{2}\phi^{'}\frac{\xi^{2}}{k^{2}_{F}}+..) \delta(\epsilon_{{\bf k}\sigma} -\epsilon_{{\bf k}^{'}\sigma}).
\end{array}
\end{eqnarray}
We keep the approximation that the dimensionless mistracking is
small such that $\frac{k_{x}\xi^{4}k^{'}_{x}}{k^{4}_{F}}\ll1$, and
write the integral as ;
\begin{equation}
I_{1} =\frac{2\pi c_{i}}{\hbar}\displaystyle\int
d^{'}kk^{'2}\delta(\epsilon_{{\bf k}\sigma}-\epsilon_{{\bf
k'}\sigma'})d^{'}\theta
d^{'}\phi^{'}\frac{\xi^{2}}{k^{2}_{F}}\\\left(A^{2}k^{'2}\sin^{3}\theta^{'}\cos^{2}\phi^{'}+2AB\sin^{2}\theta^{'}\cos\phi^{'}+B^{2}\sin\theta^{'}\right)
\end{equation}
We now evaluate the integrals over k-space angle, as these are known
analytically, as follows :

\begin{eqnarray}
I_{1} =\frac{2\pi c_{i}}{\hbar}\displaystyle\int
d^{'}kk^{'2}\delta(\epsilon_{{\bf k}\sigma}-\epsilon_{{\bf
k'}\sigma'})d^{'}\theta
d^{'}\phi^{'}\frac{\xi^{2}}{k^{2}_{F}}\left(A^{2}k^{'2}\frac{4\pi}{3}+2AB\pi.0+B^{2}.0\right)
\end{eqnarray}
which gives the result for $I_{1}$  and this becomes, upon substitution for A
;

\begin{equation}
I_{1} = \frac{k^{4}\xi^{2}4\pi(v+j)^2}{3k^{2}_{F}},
\end{equation}
to order $\xi^{4}$. The integral of the term $\delta(\epsilon_{{\bf
k}\sigma}-\epsilon_{{\bf k'}\sigma'})$ is constant with k-space
polar angles as the band energy in the simplest case depends only on
$|{\bf k}|^{2}$ and there is only a non-zero contribution to the
integral for $\delta(\epsilon_{{\bf k}\sigma}-  \epsilon_{{\bf
k'}\sigma'})= \delta(\beta (k^{'2}-k^{2})+(\sigma-\sigma^{'})J)$, in
the case where we have ${\bf k}^{'}$ =${\bf k}$ and
$\sigma^{'}=\sigma$. We turn our attention now to the integral
$I_{2}$, which can be written as :

\begin{eqnarray}
\begin{array}{c}
  I_{2} =\frac{2\pi c_{i}}{\hbar}\displaystyle\int
dk^{'}k^{'2}\delta(\epsilon_{{\bf k}\sigma}-\epsilon_{{\bf
k'}\sigma'})d\theta^{'}.\\
  d\phi^{'}\left(C^{2}\sin\theta^{'}+2CD\frac{\sin^{2}\theta^{'}\cos\phi^{'}k^{'}_{x}\xi^{2}}{k^{2}_{F}}+\frac{k^{'2}\sin^{3}\theta^{'}\cos^{2}\phi^{'}D^{2}\xi^{4}}{k^{2}_{F}}\right)\left(1-2\frac{(k_{x}k^{'}_{x})^{2}}{k^{2}_{F}}\xi^{2}+..\right)
\end{array}
\end{eqnarray}

\begin{equation}
\begin{array}{c}
  I_{2} =\frac{2\pi c_{i}}{\hbar}\displaystyle\int
dk^{'}k^{'2}\delta(\epsilon_{{\bf k}\sigma}-\epsilon_{{\bf
k'}\sigma'})d\theta^{'} d\phi^{'} \\
  .\left(C^{2}\sin\theta^{'}-C^{2}\sin^{3}\theta^{'}\cos^{2}\phi^{'}\frac{2(k_{x}k^{'}_{x})^{2}}{k^{2}_{F}}\xi^{2}
+2CD\frac{\sin^{2}\theta^{'}\cos\phi^{'}k^{'}_{x}\xi^{2}}{k^{2}_{F}}\right)\\
-\left(\frac{k^{'2}\sin^{3}\theta^{'}\cos^{2}\phi^{'}C^{2}\xi^{4}}{k^{2}_{F}}-4k^{'3}DC\sin^{4}\theta^{'}\cos^{3}\phi^{'}k^{2}_{x}\frac{\xi}{k^{4}_{F}}\right)
\end{array}
\end{equation}

Evaluating these integrals over k space spherical polar angles, we
now have :

\begin{equation}
\begin{array}{c}
  I_{2} =\frac{2\pi c_{i}}{\hbar}\displaystyle\int
dk^{'}k^{'2}\delta(\epsilon_{{\bf k}\sigma}-\epsilon_{{\bf
k'}\sigma'})d\theta^{'} d\phi^{'} \\
  .\left(C^{2}.0-C^{2}\frac{4\pi}{3}\frac{2(k_{x}k^{'}_{x})^{2}}{k^{2}_{F}}\xi^{2}
+2CD\frac{\frac{ \pi
.0}{2}k^{'}_{x}\xi^{2}}{k^{2}_{F}}-\frac{k^{'2}\frac{4\pi}{3}C^{2}\xi^{4}}{k^{2}_{F}}-4k^{'3}DC\frac{3
\pi.0}{8}k^{2}_{x}\frac{\xi}{k^{4}_{F}}\right)
\end{array}
\end{equation}
Integrating over k$^{'}$ and substituting in the definitions for C
and D, we have :

\begin{eqnarray}
\frac{2\pi c_{i}}{\hbar} k^{2}\left(-\frac{(v+\sigma j)^{2}8\pi
\xi^{2} k^{2}k^{2}_{x}}{3 k^{2}_{F}}\right) = -k^{4}\frac{2\pi
c_{i}}{\hbar}(v+\sigma
j)^2\frac{8\pi}{3}\frac{k^{2}_{x}\xi^{2}}{k^{2}_{F}}
\end{eqnarray}
We are now in a position to write down the total spin dependent
scattering time as
\begin{equation}
[\tau^{\sigma}(\overrightarrow{k})]^{-1}=\frac{2\pi
c_{i}}{\hbar}\left( \frac{k^{4}\xi^2 4 \pi (v+j)^2}{3 k^{2}_{F}} -
\frac{k^{2}k^{2}_{x}\xi^{2}8\pi (v+\sigma j)^{2}}{3
k^{2}_{F}}\right) \label{tau}
\end{equation}
The equation above defines the momentum scattering time for the spin
channel $\sigma$=$\pm$ which refers to pure spins states. This is
now used to solve the Blotzmann equation for the non-equilibrium
distribution of electronic momentum which gives rise to the spin
dependent diffusive current.
\section{\label{sec:level4} Analytical expression for DW conductivity}

We start by finding the appropriate distribution function for the electrons
in the metal, by writing down the first order solution to the Boltzmann equation as
$f^({\bf k},t)=f^{\sigma} ({\bf k}-\frac{e{\bf E}}{\hbar}t)$. This is the distribution function for electrons
in a field and the rate of change of this distribution function is given by :

\begin{equation}
\left(\frac{\partial f}{\partial t}\right)_{field} = \frac{\partial{\bf k}}{\partial t}.\nabla_{{\bf k}}f^{\sigma}({\bf k})=\frac{-e{\bf E}}{\hbar}.\nabla_{{\bf k}} f^{\sigma}(\epsilon_{{\bf k}})
\label{Boltzmann_1}
\end{equation}
To first order in the electric field ({\bf E}, with the convention e$<$0), we can write $f^{\sigma}({\bf k}) = f_{0}(\epsilon_{{\bf k}\sigma})$, which simply corresponds to the Fermi-Dirac distribution and we now expand the last term in the Equation \ref{Boltzmann_1} above, as follows ;
\begin{eqnarray}
\nabla_{{\bf k}} f^{\sigma}(\epsilon_{{\bf k}}) = \frac{d f_{0}}{d\epsilon_{{\bf k}\sigma}}\nabla\epsilon_{{\bf k}\sigma}\\
=-\hbar {\bf v}^{\sigma}_{{\bf k}\sigma}\delta(\epsilon_{F}-\epsilon_{{\bf k}\sigma}),
\end{eqnarray}
which gives us the field term in the distribution function rate equation :
\begin{equation}
\left(\frac{\partial f}{\partial t}\right)_{field} = +e{\bf v}^{\sigma}_{\bf k}.{\bf E}\delta(\epsilon_{F}-\epsilon_{{\bf k}\sigma}).
\end{equation}
We now turn our attention to the collision terms in the first order time derivative of the distribution function, and we begin with the spinless version :
\begin{equation}
\left(\frac{\partial f}{\partial t}\right)_{coll} = \displaystyle\sum_{{\bf k}^{'}}\left(W_{{\bf k}^{'}{\bf k}}f({\bf k}^{'})[1-f({\bf k})]-W_{{\bf k}{\bf k}^{'}}f({\bf k})[1-f({\bf k}^{'})]\right).
\end{equation}
where $W_{{\bf k}^{'}{\bf k}}$ are the scattering rates (in units of energy per unit time).
The first term in the equation above represents the 'scattering in' terms, while the second represents the
'scattering out' terms and for elastic scattering, we have $W_{{\bf k}{\bf k}^{'}}= W_{{\bf k}^{'}{\bf k}}$ which arises due to the time-reversal symmetry which inelastic processes obey.

\begin{eqnarray}
\left(\frac{\partial f}{\partial t}\right)_{coll} = \displaystyle\sum_{{\bf k}^{'}}W_{{\bf k}{\bf k}^{'}}[f({\bf k}^{'})-f({\bf k})]\\
\frac{\Omega}{8\pi^{3}}\displaystyle\int d^{3}{\bf k}^{'}W_{{\bf k}{\bf k}^{'}}[f({\bf k}^{'}) -f({\bf k})]
\end{eqnarray}

Now we recast these equations in spin-dependent form, and now write the collision term, as follows ;
\begin{eqnarray}
\left(\frac{\partial f^{\sigma}}{\partial t}\right)_{coll} = \frac{\Omega}{8\pi^{3}}\displaystyle\int d^{3}{\bf k}^{'}\left(W^{\sigma\sigma}_{{\bf k}{\bf k}^{'}}[f^{\sigma}({\bf k}^{'})-f^{\sigma}({\bf k})]+W^{\sigma-\sigma}_{{\bf k}{\bf k}^{'}} [f^{-\sigma}({\bf k}^{'})-f^{\sigma}({\bf k})]\right)
\end{eqnarray}
and we invoke the steady-state condition :
\begin{equation}
\left(\frac{\partial f^{\sigma}}{\partial t}\right)_{field}+\left(\frac{\partial f^{\sigma}}{\partial t}\right)_{coll}=0.
\end{equation}
We now arrive at the appropriate Boltzmann equation for the spin dependent transport of electrons in the metal, in the most general sense :

\begin{equation}
+e{\bf v}^{\sigma}_{{\bf k}}.{\bf E}\delta(\epsilon_{F} - \epsilon_{{\bf k}\sigma})=\frac{\Omega}{8\pi^{3}}\displaystyle\int d^{3}{\bf k}\left(W^{\sigma\sigma}_{{\bf k}^{'}k}[f^{\sigma}({\bf k})-f^{\sigma}({\bf k'})]+W^{\sigma-\sigma}_{{\bf k}{\bf k}^{'}}[f^{\sigma}({\bf k})-f^{-\sigma}({\bf k'})]\right)
\end{equation}

\begin{eqnarray}
+e{\bf v}^{\sigma}_{{\bf k}}.{\bf E}\delta(\epsilon_{F} - \epsilon_{{\bf k}\sigma})=\frac{\Omega}{8\pi^{3}}\frac{2\pi}{\hbar}\displaystyle\int d^{3}{\bf k}|V^{\sigma\sigma}_{{\bf k}^{'}k}|^{2}[f^{\sigma}({\bf k})-f^{\sigma}({\bf k'})]\delta(\epsilon_{{\bf k}\sigma}-\epsilon_{{\bf k}^{'}\sigma})+\\
|V^{\sigma-\sigma}_{{\bf k}{\bf k}^{'}}|^{2}[f^{\sigma}({\bf k})-f^{-\sigma}({\bf k'})]\delta(\epsilon_{{\bf k}\sigma}-\epsilon_{{\bf k}^{'}-\sigma})
\end{eqnarray}
Now, for a two-band ferromagnet, we have the following relations for intra-spin-band and inter-spin-band scattering, respectively, as follows :
\begin{eqnarray}
\epsilon_{{\bf k}\sigma}-\epsilon_{{\bf k}^{'}\sigma} = \frac{\hbar^{2}}{2m^{*}}\left(k^{2}-k^{'2}\right)\\
\epsilon_{{\bf k}\sigma}-\epsilon_{{\bf k}^{'}-\sigma} = \frac{\hbar^{2}}{2m^{*}}\left(k^{2}-k^{'2}\right)+2\sigma J
\end{eqnarray}
We also expand the Boltzmann distribution about the Fermi energy and write the non-equilibrium distribution function for the electrons as follows :
\begin{equation}
f^{\sigma}({\bf k}) = f_{0}(\epsilon_{{\bf k}\sigma}) + n^{\sigma}({\bf k})\delta(\epsilon_{F}-\epsilon_{{\bf k}\sigma})
\end{equation}

\begin{eqnarray}
\frac{\Omega}{4\pi^{2}}\displaystyle\int d^{3}{\bf k}|V^{\sigma\sigma}_{{\bf k}^{'}k}|^{2}[n^{\sigma}({\bf k})\delta(\epsilon_{F}-\epsilon_{{\bf k}^{'}\sigma})-n^{\sigma}({\bf k^{'}})\delta(\epsilon_{F}-\epsilon_{{\bf k}^{'}\sigma})]\delta(\epsilon_{{\bf k}\sigma}-\epsilon_{{\bf k}^{'}\sigma})+\\
|V^{\sigma-\sigma}_{{\bf k}{\bf k}^{'}}|^{2}[n^{\sigma}({\bf k})\delta(\epsilon_{F}-\epsilon_{{\bf k}^{'}\sigma})-n^{-\sigma}({\bf k^{'}})\delta(\epsilon_{F}-\epsilon_{{\bf k}^{'}-\sigma})]\delta(\epsilon_{{\bf k}\sigma}-\epsilon_{{\bf k}^{'}-\sigma})
\end{eqnarray}

\begin{eqnarray}
+e{\bf v}^{\sigma}_{{\bf k}}.{\bf E}=\frac{\Omega}{4\pi^{2}}\displaystyle\int d^{3}{\bf k}|V^{\sigma\sigma}_{{\bf k}^{'}k}|^{2}[n^{\sigma}({\bf k})-n^{\sigma}({\bf k^{'}})]\delta(\epsilon_{F}-\epsilon_{{\bf k}^{'}\sigma})+\\
|V^{\sigma-\sigma}_{{\bf k}{\bf k}^{'}}|^{2}[n^{\sigma}({\bf k})-n^{-\sigma}({\bf k^{'}})]\delta(\epsilon_{F}-\epsilon_{{\bf k}^{'}-\sigma})
\end{eqnarray}

For the conductivity in the "current in wall" geometry, we solve the
Boltzmann equation with the following distribution function :
\begin{equation}
f^{\sigma}({\bf k})=f^{o}({\bf k})-e
v^{\sigma}_{y}\delta(\epsilon_{F}-\epsilon_{{\bf
k}\sigma})\tau^{\sigma}({\bf k})\\
\end{equation}
 and write the total conductivity $\sigma_{CIW} = \displaystyle\sum_{\sigma}\int-e f^{\sigma}({\bf k}){\bf v_{\sigma}}d^{3}{\bf k}$ of the system as
 \begin{equation}
\sigma_{CIW} = \displaystyle\int d^{3}{\bf
k}\left(-e^{2}v^{\uparrow}_{y}E\delta(E_{F}-E_{{\bf
k}\uparrow})\tau^{\uparrow}({\bf
k})-e^{2}v^{\downarrow}_{y}\delta(E_{F}-E_{{\bf
k}\downarrow})\tau^{\downarrow}({\bf k}) + f^{o}({\bf
k})e(v^{\uparrow}_{y}+v^{\downarrow}_{y})\right)
\end{equation}
The first two terms in the above equation are the spin dependant
conductivities $\sigma_{\uparrow}$ and $\sigma_{\downarrow}$
respectively, while the last term is just the total conductivity of
the two spin channels without spin relaxation.\\
Takin the expressions for the total scattering rates from Equation
\ref{scattering_rates}, we set the scattering in momentum component
k$_{x}$ to zero (the current is 'in wall') and write the the total
spin relaxation time as :
\begin{eqnarray*}
\begin{array}{c}
  [\tau({\bf k})]^{-1}=\frac{2\pi
c_{i}}{(2\pi)^{3}\hbar}\displaystyle\int
dk^{'3}N^{2}(k_{x})\delta(\epsilon_{k\sigma}-\epsilon_{k^{'}\sigma})\left[(v+\sigma
j)^{2}+(v+j)^2k^{'2}_{x}\frac{\xi^{2}}{k^{2}_{F}}\right] \\
  =\frac{c_{i}}{4\pi^{2}\hbar}\displaystyle\int d
k^{'}k^{'2}\delta(\epsilon_{k\sigma}-\epsilon_{k^{'}\sigma})\\
\left(\sin\theta^{'}(v+\sigma j)^{2}-(v+\sigma
j)^{2}k^{'2}\sin^{3}\theta^{'}\cos^{2}\phi^{'}\frac{\xi^{2}}{k^{2}_{F}}+(v+j)^{2}k^{'2}\sin^{3}\theta^{'}\cos^{2}\phi^{'}\frac{\xi^{2}}{k^{2}_{F}}\right)d
\theta^{'}d\phi^{'} \\
  =\frac{c_{i}}{4\pi^{2}\hbar}\displaystyle\int d
k^{'}k^{'2}\delta(\epsilon_{k\sigma}-\epsilon_{k^{'}\sigma})\left(4\pi
(v+\sigma j)^{2}-(v+\sigma j)^{2}k^{'2}\frac{4
\pi}{3}\frac{\xi^{2}}{k^{2}_{F}}+(v+j)^{2}k^{'2}\frac{4\pi}{3}\frac{\xi^{2}}{k^{2}_{F}}\right)\\
=\frac{c_{i}}{4\pi^{2}\hbar}\left(k^{2}(4\pi(v+\sigma j)^2-(v+\sigma
j)^{2}k^{2}\frac{4
\pi}{3}\frac{\epsilon^{2}}{k^{2}_{F}}+(v+j)^{2}\frac{k^{2}\epsilon^{2}4\pi}{3
k^{2}_{F}})\right).
\end{array}
\end{eqnarray*}
We now solve the Boltzmann equation, which is written as follows :
\begin{equation}
-e {\bf v}^{\sigma}.{\bf E}\delta(\epsilon_{F}-\epsilon_{{\bf
k}\sigma})=\frac{1}{8 \pi^{3}}\displaystyle\int
W^{\sigma\sigma}_{\bf k k^{'}}\left[f^{\sigma}({\bf
k})-f^{\sigma}({\bf k^{'}})\right]d^{3}{\bf k^{'}}+\frac{1}{8
\pi^{3}}\displaystyle\int W^{\sigma-\sigma}_{\bf k
k^{'}}\left[f^{\sigma}({\bf k})-f^{-\sigma}({\bf
k^{'}})\right]d^{3}{\bf k^{'}}
\end{equation}
which, for the geometry under consideration, we write the solution
to the Boltzmann equation as $f^{\sigma}({\bf k})=f_{0}({\bf
k})-ev^{\sigma}_{y}E\delta(\epsilon_{F}-\epsilon_{{\bf
k}\sigma})\tau^{\sigma}({\bf k})$, which is justified as integration
over ${\bf k^{'}}$ results in the vanishing of scattering in ${\bf k}$
components. We use this solution to write the conductivity of the
wall with the current flowing 'in wall' $\sigma_{CIW}$.\\
\begin{equation}
\sigma_{CIW}=\sigma_{0} +
\displaystyle\sum_{\sigma}\int\frac{-e^{2}\frac{\hbar^{2}}{m^{*2}}k_{\sigma}^{2}\sin^{2}\theta\sin^{2}\phi
E\delta(\epsilon_{F}-\epsilon_{{\bf k} \sigma})\sin\theta k^{2}d k
d\theta d\phi}{\frac{c_{i}}{4\pi^{2}\hbar}\left(4\pi k^{2}(v+\sigma
j)^{2}-(v+\sigma j)^{2}\frac{4\pi k^{4}\xi^{2}}{3
k^{2}_{F}}+(v+j)^{2}\frac{k^{4}\xi^{2}4\pi}{3 k^{2}_{F}}\right)}\\
\end{equation}
We now find that the total conductivity can be written as:
\begin{equation}
\begin{array}{c}
   \sigma_{CIW} = \sigma_{0}-\frac{4\pi e^{2}
\frac{\hbar^{2}}{m^{*2}}}{\frac{c_{i}}{4\pi^{2}\hbar}}\left(\frac{k^{\uparrow
2}_{F}}{4\pi k^{\uparrow 2}_{F}(v+j)^2-(v+j)^2 \frac{4\pi
k^{\uparrow 4}_{F}\xi^{2}}{k^{2}_{F}}+(v+j)^2 \frac{4\pi k^{\uparrow
4}_{F}\xi^{2}}{k^{2}_{F}}}
    +\frac{k^{\downarrow 2}_{F}}{4\pi k^{\downarrow 2}_{F}(v-j)^2-(v-j)^2 \frac{4\pi
k^{\downarrow 4}_{F}\xi^{2}}{k^{2}_{F}}+(v+j)^2 \frac{4\pi
k^{\downarrow 4}_{F}\xi^{2}}{k^{2}_{F}}}\right)
\end{array}
\end{equation}

\begin{eqnarray}
  \sigma_{CIW} = \sigma_{0}-\frac{4\pi e^{2}
\frac{\hbar^{2}}{m^{*}}}{\frac{c_{i}}{4\pi^{2}\hbar}}\left(\frac{k^{\uparrow
2}_{F}}{4\pi k^{\uparrow 2}_{F}(v+j)^2}
  +\frac{k^{\downarrow 2}_{F}}{4\pi k^{\downarrow 2}_{F}(v-j)^2-(v-j)^2 \frac{4\pi
k^{\downarrow 4}_{F}\xi^{2}}{k^{2}_{F}}+(v+j)^2 \frac{4\pi
k^{\downarrow 4}_{F}\xi^{2}}{k^{2}_{F}}}\right)
\end{eqnarray}

For the case for non-vanishing k$_{x}$, where the current has a
component perpendicular to the wall. Inserting the formulae for the
spin scattering (relaxation) lifetimes from equation \ref{tau}, we
get :

\begin{equation}
\sigma_{\uparrow} = \frac{-e^{2}2\pi
c_{i}}{\hbar}\displaystyle\int\left(\frac{\hbar
k^{\uparrow}_{y}}{m^{*}}\right)^{2}\frac{\delta(E_{F}-E_{{\bf
k}\uparrow})d^{3}{\bf
k}}{\frac{k^{4}}{k^{2}_{F}}\xi^{2}\frac{4\pi}{3}(v+j)^{2}-\frac{k^{2}}{k^{2}_{F}}\xi^{2}\frac{8\pi}{3}(v+j)^2k^{2}_{x}}
\end{equation}
\begin{eqnarray}
=\frac{-e^{2}\hbar 2\pi c_{i}}{m^{*2}}\displaystyle\int\left(
k^{\uparrow}_{y}\right)^{2}\frac{\sin^{3}\theta\sin^{2}\phi d\phi
d\theta\delta(E_{F}-E_{{\bf k}\uparrow})k^{2}d
k}{\frac{k^{4}4\pi(v+j)^{2}\xi^{2}}{3
k^{2}_{F}}\left(1-\frac{4}{2}\sin^{2}\theta\cos^{2}\phi\right)}
\end{eqnarray}
\begin{eqnarray}
\begin{array}{c}
  \sigma_{\downarrow} = \frac{-e^{2}\hbar 2\pi
c_{i}}{m^{*2}}\displaystyle\int\left(
k^{\downarrow}_{y}\right)^{2}\frac{\sin^{3}\theta\sin^{2}\phi d\phi
d\theta\delta(E_{F}-E_{{\bf k}\downarrow})k^{2}d
k}{\frac{k^{4}4\pi\xi^{2}}{3
k^{2}_{F}}\left[(v+j)^2-\frac{4}{2}(v-j)^{2}\sin^{2}\theta\cos^{2}\phi\right]} \\
= \frac{-e^{2}\hbar 2\pi c_{i}}{m^{*2}}\displaystyle\int
k^{4}\frac{\sin^{3}\theta\sin^{2}\phi d\phi
d\theta\delta(E_{F}-E_{{\bf k}\downarrow})k^{2}d
k}{\frac{k^{4}4\pi\xi^{2}}{3
k^{2}_{F}}\left[(v+j)^2-\frac{4}{2}(v-j)^{2}\sin^{2}\theta\cos^{2}\phi\right]}  \\
  = \frac{-e^{2}\hbar 2\pi c_{i}}{m^{*2}}\displaystyle\int
k^{4}\frac{\sin^{3}\theta\sin^{2}\phi d\phi
d\theta\delta(E_{F}-E_{{\bf k}\downarrow})k^{2}d
k}{\frac{k^{4}4\pi\xi^{2}}{3
k^{2}_{F}}\left[(v+j)^2-\frac{4}{2}(v-j)^{2}\sin^{2}\theta\cos^{2}\phi\right]}
\end{array}
\end{eqnarray}
As as result of the integration over ${\bf k}$, we find that the
expressions for the spin dependent conductivity in the presence of a
domain wall are as follows :
\begin{eqnarray}
\sigma_{\uparrow} =\frac{-e^{2}\hbar 2\pi c_{i}}{m^{*2}}k^{2}_{F}
\frac{3
}{4\pi\xi^{2}}\displaystyle\int\frac{\sin^{3}\theta\sin^{2}\phi
d\phi d \theta}{(v+j)^{2}(1-2\sin^{2}\theta\cos^{2}\phi)}\\
\sigma_{\downarrow} =\frac{-e^{2}\hbar 2\pi c_{i}}{m^{*2}}k^{2}_{F}
\frac{3
}{4\pi\xi^{2}}\displaystyle\int\frac{\sin^{3}\theta\sin^{2}\phi
d\phi d \theta}{(v+j)^{2}1-4\sin^{2}\theta\cos^{2}\phi (v-j)^{2}}\\
\label{sigma_relations}
\end{eqnarray}
Thus, the contribution to the resistivity is positive, as we have
$\sigma_{CIW}=\sigma_{0}+\sigma_{\uparrow}+\sigma_{\downarrow}$ and
$\rho=\sigma^{-1}$.  The $\sigma_{\uparrow}$ integral over angles in
${\bf k}$ space can be evaluated above to give the result:

\begin{equation}
\sigma_{\uparrow} =\frac{-e^{2}\hbar 2\pi c_{i}}{m^{*2}}k^{2}_{F}
\frac{3}{4\pi\xi^{2}(v+j)^{2}} 11.6071,
\end{equation}
while the $\sigma_{\downarrow}$ cannot be integrated directly
without prior knowledge of v and j. The current perpendicular to
wall geometry (${\bf k} \bot \nabla\theta$) can be solved. Let us
now turn to the problem of the ratio of the conductivity of the spin
channels, which is given by $\alpha$
=$\sigma_{\uparrow}/\sigma_{\downarrow}$, which we can write 
using Equations \ref{sigma_relations} as :

\begin{equation}
\alpha  = \frac{\displaystyle\int\frac{\sin^{3}\theta\sin^{2}\phi
d\phi d\theta}{(v+j)^{2}(1-2
\sin^{2}\cos^{2}\phi)}}{\displaystyle\int\frac{\sin^{3}\theta\sin^{2}\phi
d \theta d \phi}{(v+j)^{2}-4\sin^{2}\cos^{2}\phi(v-j)^{2}}}.
\end{equation}

\begin{figure}[ht]
\begin{center}
 \includegraphics[width=4.0in]{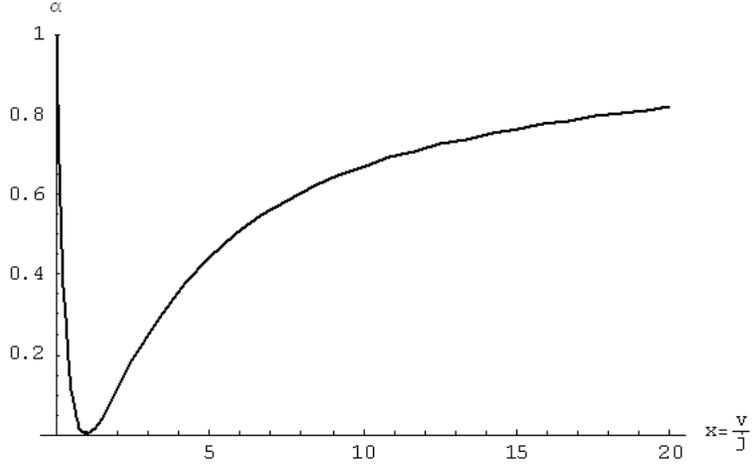}\\
\end{center}
 \caption{(Color Online) Plot of $\alpha$ Vs. the ratio of the impurity coulomb scattering
 potential and the exchange coupling to the impurity.  } \label{alpha_plot}
\end{figure}

These integrals can be evaluated analytically, as follows :
\begin{eqnarray}
\begin{array}{c}
\frac{\sigma_{\uparrow}}{\sigma_{\downarrow}}
=\frac{\displaystyle\int_{0}^{\pi}(\frac{\phi}{2}
cosech^{2}\theta-\frac{1}{2}tan^{-1}(\frac{\tan\phi}{\sqrt{\cos^{2}\theta}})\sqrt{\cos(2
\theta)}cosec^{2}(\theta)\sin^{3}\theta)\|_{0}^{2\pi}}{}\\
{\displaystyle\int_{0}^{\pi} \frac{\phi
 cosech^{2}\theta}{4
(j-v)^{2}}+\frac{tanh^{-1}(j^{2}+2vj+v^{2})\tan\phi}{\sqrt{j^{4}-4
j^{3}v^{2}-4j v^{3}+v^{4}-2 j^{4}\cos2\theta+4 j^{2}v^{2}\cos
2\theta-2v^{4}\cos2\theta}}  +}\\
{\sin^{3}\theta \frac{(-j^{2}+6 j v-v^{2}+2
j^{2}\cos2\theta-4jv\cos2\theta+2 v^{2}\cos 2\theta
cosech^{2}\theta)}{4(j-v)^{2}\sqrt{j^4-4j^3v-10j^{2}v^{2}-4
jv^{3}+v^{4}-2j^{4}\cos2\theta+4j^{2}v^{2}\cos2\theta+
4j^{2}v^{2}\cos2\theta-2v^{4}\cos2\theta}}\|_{0}^{2\pi}}
\end{array}
\end{eqnarray}
Evaluating the right hand side of the above equation, we arrive at
the simple relation :

\begin{equation}
\frac{\sigma_{\uparrow}}{\sigma_{\downarrow}} =
\frac{\displaystyle\int_{0}^{\pi}(v-j)^{2}\pi\sin\theta
d\theta}{\displaystyle\int_{0}^{\pi}\pi \sin\theta d\theta
(v+j)^{2}}=\left(\frac{v-j}{v+j}\right)^{2}
\end{equation}
This establishes the dependency of the spin dependent conductivity
asymmetry parameter $\alpha$ on the strength of the impurity
potential v and the exchange coupling (j) to the impurity. When there is an asymmetry in the impurity and
exchange strengths (v$\gg$j), we asymptotically approach the
completely unpolarized current $\frac{\alpha -
1}{\alpha+1}\longrightarrow$0. Similarly, when the exchange
potential strength vanishes (j$\longrightarrow$0), we also tends
towards the unpolarized case. On the contrary, when j=v, we now have
a completely spin polarized case. The results of this calculation
are plotted in Figure \ref{alpha_plot}

\newpage


\begin{acknowledgments}

\end{acknowledgments}

\appendix
\section{Perturbation Expansion to second order}
We begin with a Hamiltonian which characterizes the ground state
wavefunctions of the the total wavefunction $\Psi\rangle$  of the
system . The Hamiltonian eigenvalue problem is written as follows :
\begin{equation}
H|\Psi\rangle=E|\Psi\rangle
\end{equation}
where $|\Psi\rangle$ is now written as an expansion of the ground
states kets, which is linear superposition of orthogonal functions
$|n^{(0)}\rangle$, $|k^{(1)}\rangle$, $|m^{(2)}\rangle$,... as
follows

\begin{equation}
(H_{0}+V)(|n^{0}\rangle \lambda |k^{1}\rangle+\lambda^{2}
|m^{2}\rangle+...)=(E^{(0)}+\lambda
E^{(2)}+E^{(2)}\lambda^{2}+....)( |n^{0}\rangle \lambda
|k^{1}\rangle+\lambda^{2} |m^{2}\rangle+...) \label{A1}
\end{equation}
where the superindex $^{n}$ refers to the order of the expansion.
Since they are coefficients of an order-n polynomial, they are
linearly independent.

\begin{figure}[ht]
\begin{center}
 \includegraphics[width=3.0in]{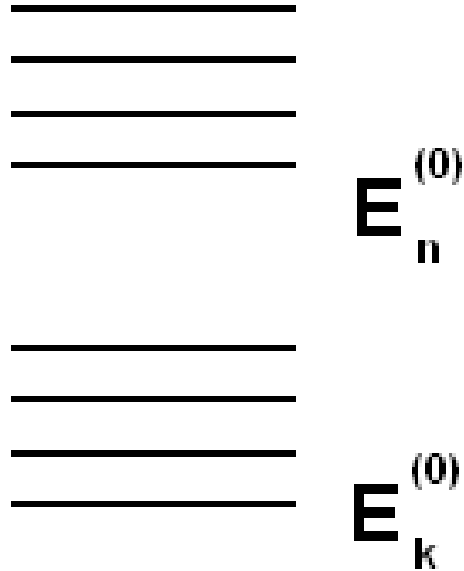}\\
\end{center}
 \caption{(Color Online) Spectrum of energy levels in a degenerate system.} \label{FigA2_}
\end{figure}

Taking the O(1) equation from \ref{A1} above, we have :

\begin{eqnarray}
O(1) : H_{0} |n^{0}\rangle =E_{0} |n^{0}\rangle\\
\Rightarrow\\
 \langle n^{0}|H_{0} |n^{0}\rangle =E_{0}\langle n^{0}|n^{0}\rangle\\
O(\lambda) : \lambda H_{0} |k^{1}\rangle+\lambda V |n^{0}\rangle =\lambda E_{0} |k^{1}\rangle+\lambda E^{1} |n^{0}\rangle\\
O(\lambda^{2}) : \lambda^{2} H_{0} |m^{2}\rangle+\lambda^{2} V |k^{1}\rangle =\lambda^{2} E_{1} |k^{1}\rangle+\lambda^{2} E^{2} |n^{0}\rangle\\
\end{eqnarray}
We now arrive at the expansion coefficients for the first order
correction to the wavefunctions , by multiplying the O($\lambda$)
equation above by the  $\langle n^{0}|$ ket and we arrive at the
following equation :

\begin{equation}
\lambda\langle n^{0}|H_{0}| k^{1}\rangle +\lambda\langle n^{0}|V|
n^{0}\rangle= \lambda E^{(0)}_{n}\langle n^{(0)}|k^{(1)}\rangle
+\lambda E^{(1)}\langle n^{(0)}|n^{(0)}\rangle
\end{equation}
$\vee$ n $\neq$ k, we have :
\begin{equation}
\langle n^{(0)}|V|n^{(0)}\rangle=E^{(1)},
\end{equation}
which is the first-order correction to the total energy. Taking the
O($\lambda$) equation, we multiply across by the ket
$|k^{(0)}\rangle$ armed with the decomposition of $|k^{(1)}\rangle$
onto the vector space $|k^{(0)}\rangle$, which we can write as
$|k^{(1)}\rangle$ = $\sum_{k} C_{kn}$ $|k^{(0)}\rangle$. This can be
written more succinctly in the outer product notation
$|k^{(1)}\rangle$ = $|k^{(1)}\rangle \langle
n^{(0)}|n^{(0)}\rangle$. We now write
the O($\lambda$) equation as :\\
\begin{eqnarray}
\lambda \langle|k^{(0)}|H_{0}\|k^{(1)}\rangle+\lambda
\langle|k^{(0)}|V|n^{(0)}\rangle=\lambda E^{(0)}_n\langle
k^{(0)}|k^{(1)}\rangle+\lambda E^{(1)}\langle
k^{(0)}|n^{(0)}\rangle.
\end{eqnarray}
Now, $\vee$ n $\neq$ k, we seek the coefficients of the expansion of
the first order wavefunction in ground state kets, as follows : \\
\begin{eqnarray}
\displaystyle\sum_{n\neq k}\langle
k^{(0)}|H_{0}C_{k}|k^{(0)}\rangle+\langle
k^{(0)}|V|n^{(0)}\rangle=E^{(0)}_{n}\langle k^{(0)}
C_{k}|k^{(0)}\rangle+E^{(1)}\langle k^{(0)}|n^{(0)}\rangle\\
\Rightarrow\\
C_{k}^{(1)} = \displaystyle\sum_{n\neq k}
\frac{|V_{kn}|}{E^{0}_{n}-E^{(0)}_{k}}
\end{eqnarray}
In order to find the second order perturbation expansion
coefficients, we iterative this progress further by using the O
($\lambda^{2}$), which has the following form :
\begin{eqnarray}
H_{0}|m^{(2)}\rangle
+V|k^{(1)}\rangle=E^{(0)}|m^{(2)}\rangle+E^{(1)}|k^{(1)}\rangle+E^{(2)}|n^{(0)}\rangle\\
H_{0}C_{m}|m^{(0)}\rangle+V
C_{k}|k^{(0)}\rangle=E^{(0)}C_{m}|m^{(0)}\rangle +E^{(1)}C_{k}
|k^{(0)}\rangle + E_{n}^{(2)} |n^{(0)}\rangle\\
(E^{(0)}_{m}-E^{(0)}_{n})C_{m} |m^{(0)}=V
C_{k}|k^{(0)}+E^{(1)}_{k}C_{k} |k^{(0)}\rangle +
E^{(2)}_{m}|n^{(0)}\rangle
\end{eqnarray}
Next, we multiply across by the ket $|m^{0}\rangle$ and make use of
the fact that the each basis $|n^{0}\rangle$,
$|k^{0}\rangle$,$|m^{0}\rangle$ are all orthonormal sets, as follows
:
\begin{equation}
C_{m}(E^{(0)}_{m}-E^{(0)}_{n})=V_{mk}C_{k}+E^{(1)}_{k}
C_{k}\delta_{mk} +E^{(2)}_{m}\delta_{mn}
\end{equation}
Next we take the case, whereby n$\neq$k$\neq$m, which gives :
\begin{equation}
C_{m} = \displaystyle\sum_{m\neq n} \frac{V_{m k}
C_{k}}{E^{(0)}_{m}-E^{(0)}_{n}} = \displaystyle\sum_{m\neq
n}\displaystyle\sum_{n\neq k} \frac{V_{mk}
V_{kn}}{(E^{(0)}_{m}-E^{(0)}_{n})(E^{(0)}_{n}-E^{(0)}_{k})}.
\end{equation}
In the above equation, we have substituted in the result for the
first order expansion coefficients C$_{k}$. The second order
corrections to the total wavefunction are now know in terms of the
matrix elements of the perturbing potential V and the ground state
eigenenergies. To find the second order corrections to the energy
$E^{(2)}_{n}$, take n=m$\neq$k in the above equation, which gives :
\begin{equation}
E^{(2)}_{n}= -V_{m m}C_{k}
\end{equation}

 \begin{figure}[!ht]
\begin{center}
 \includegraphics[width=1.2in]{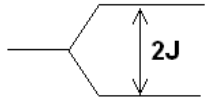}\\
\end{center}
 \caption{(Color Online) Exchange split ferromagnetic ground state.} \label{FigA2_}
\end{figure}

\bibliography{DWR-LZ-analytic}

\begin{thebibliography}{2}
\expandafter\ifx\csname natexlab\endcsname\relax\def\natexlab#1{#1}\fi
\expandafter\ifx\csname bibnamefont\endcsname\relax
  \def\bibnamefont#1{#1}\fi
\expandafter\ifx\csname bibfnamefont\endcsname\relax
  \def\bibfnamefont#1{#1}\fi
\expandafter\ifx\csname citenamefont\endcsname\relax
  \def\citenamefont#1{#1}\fi
\expandafter\ifx\csname url\endcsname\relax
  \def\url#1{\texttt{#1}}\fi
\expandafter\ifx\csname urlprefix\endcsname\relax\def\urlprefix{URL }\fi
\providecommand{\bibinfo}[2]{#2}
\providecommand{\eprint}[2][]{\url{#2}}

\bibitem[{\citenamefont{Tatara and Fukuyama}(1997)}]{PhysRevLett.78.3773}
\bibinfo{author}{\bibfnamefont{G.}~\bibnamefont{Tatara}} \bibnamefont{and}
  \bibinfo{author}{\bibfnamefont{H.}~\bibnamefont{Fukuyama}},
  \bibinfo{journal}{Phys. Rev. Lett.} \textbf{\bibinfo{volume}{78}},
  \bibinfo{pages}{3773} (\bibinfo{year}{1997}).

\bibitem[{\citenamefont{Levy and Zhang}(1997)}]{PhysRevLett.79.5110}
\bibinfo{author}{\bibfnamefont{P.~M.} \bibnamefont{Levy}} \bibnamefont{and}
  \bibinfo{author}{\bibfnamefont{S.}~\bibnamefont{Zhang}},
  \bibinfo{journal}{Phys. Rev. Lett.} \textbf{\bibinfo{volume}{79}},
  \bibinfo{pages}{5110} (\bibinfo{year}{1997}).

\end{thebibliography}

\end{document}